\documentclass[11pt,prd,onecolumn,amsmath,amssymb,aps,floats,floatfix,nofootinbib]{revtex4-2}
\usepackage[colorlinks=true,urlcolor=blue,anchorcolor=blue,citecolor=blue,filecolor=blue,linkcolor=blue,menucolor=blue,linktocpage=true]{hyperref} 

\usepackage[inline]{enumitem}
\usepackage[multidot]{grffile}  
\usepackage{dcolumn}
\usepackage{bm}
\usepackage{amsmath}
\usepackage{amsfonts}
\usepackage{amssymb}
\usepackage{color}
\usepackage[table]{xcolor}
\usepackage{float}
\usepackage{latexsym}
\usepackage{slashed} 
\usepackage{pstricks}
\usepackage{indentfirst}
\usepackage{mathrsfs}
\usepackage{multirow}
\usepackage{epsfig,psfrag}
\usepackage{graphicx}
\usepackage{enumitem}
\usepackage{geometry,amssymb,yfonts}
\usepackage{yhmath}
\usepackage{anysize}
\usepackage{subfigure}
\usepackage{mathtools}
\usepackage{setspace} 
\usepackage[utf8]{inputenc} 
\usepackage[scientific-notation=true]{siunitx} 

\usepackage{url}
\usepackage{makecell}

\graphicspath{{fig/}}

\allowdisplaybreaks 

\begin{document}

\title{Nano-Hertz gravitational waves from collapsing domain walls associated with freeze-in dark matter in light of pulsar timing array observations}

\author{Zhao Zhang$^a$}
\author{Chengfeng Cai$^{a,b}$}\email[Corresponding author. ]{caichf3@mail.sysu.edu.cn}
\author{Yu-Hang Su$^a$}
\author{Shiyu~Wang$^a$}
\author{Zhao-Huan Yu$^a$}\email[Corresponding author. ]{yuzhaoh5@mail.sysu.edu.cn}
\author{Hong-Hao Zhang$^a$}\email[Corresponding author. ]{zhh98@mail.sysu.edu.cn}
\affiliation{$^a$School of Physics, Sun Yat-Sen University, Guangzhou 510275, China}
\affiliation{$^b$School of Science, Sun Yat-Sen University, Shenzhen 518107, China}

\begin{abstract}
Evidence for a stochastic gravitational wave background in the nHz frequency band is recently reported by four pulsar timing array collaborations NANOGrav, EPTA, CPTA, and PPTA. It can be interpreted by gravitational waves from collapsing domain walls in the early universe. We assume such domain walls arising from the spontaneous breaking of a $Z_2$ symmetry in a scalar field theory, where a tiny $Z_2$-violating potential is required to make domain walls unstable. We propose that this $Z_2$-violating potential is radiatively induced by a feeble Yukawa coupling between the scalar field and a fermion field, which is also responsible for dark matter production via the freeze-in mechanism. Combining the pulsar timing array data and the observed dark matter relic density, we find that the model parameters can be narrowed down to small ranges.
\end{abstract}

\maketitle
\tableofcontents

\clearpage

\section{Introduction}

Recently, four pulsar timing array (PTA) collaborations NANOGrav~\cite{NANOGrav:2023gor, NANOGrav:2023hvm}, EPTA~\cite{Antoniadis:2023rey, Antoniadis:2023zhi}, CPTA~\cite{Xu:2023wog}, and PPTA~\cite{Reardon:2023gzh} have reported positive evidence for an isotropic, stochastic background of gravitational waves (GWs) in the nHz frequency band.
Potential sources for such a stochastic GW background (SGWB) involve supermassive black hole binaries~\cite{Ellis:2023dgf, Shen:2023pan, Ghoshal:2023fhh, Huang:2023chx, Broadhurst:2023tus, Depta:2023qst, Bi:2023tib, Westernacher-Schneider:2023cic, Gouttenoire:2023nzr, Bian:2023dnv},
first-order phase transitions~\cite{Ashoorioon:2022raz, Madge:2023cak, Han:2023olf, Megias:2023kiy, Fujikura:2023lkn, Addazi:2023jvg, Athron:2023mer, Jiang:2023qbm, Xiao:2023dbb, Li:2023bxy, Ghosh:2023aum, Athron:2023aqe, Cruz:2023lnq, Wu:2023hsa, DiBari:2023upq, Gouttenoire:2023bqy, Salvio:2023ynn, Ahmadvand:2023lpp},
cosmic strings~\cite{Ellis:2020ena, Qiu:2023wbs, Ellis:2023tsl, Wang:2023len, Kitajima:2023vre, Eichhorn:2023gat, Lazarides:2023ksx, Servant:2023mwt, Antusch:2023zjk, Buchmuller:2023aus, Yamada:2023thl},
domain walls~\cite{Chiang:2020aui, Sakharov:2021dim, King:2023cgv, Guo:2023hyp, Kitajima:2023cek, Bai:2023cqj, Blasi:2023sej, Gouttenoire:2023ftk, Barman:2023fad, Lu:2023mcz, Du:2023qvj, Li:2023tdx, Babichev:2023pbf, Gelmini:2023kvo, Ge:2023rce},
inflation~\cite{Vagnozzi:2023lwo, Borah:2023sbc, Murai:2023gkv, Datta:2023vbs, Chowdhury:2023opo, Niu:2023bsr, Unal:2023srk, Firouzjahi:2023lzg, Choudhury:2023kam, HosseiniMansoori:2023mqh, Cheung:2023ihl},
scalar-induced GWs~\cite{Cai:2023dls, Wang:2023ost, Liu:2023ymk, Figueroa:2023zhu, Yi:2023mbm, Zhu:2023faa, You:2023rmn, Balaji:2023ehk},
and other astrophysical and cosmological GW sources~\cite{Lambiase:2023pxd, Yang:2023aak, Deng:2023btv, Franciolini:2023wjm, Franciolini:2023pbf, Oikonomou:2023qfz, Inomata:2023zup, Zhang:2023lzt, Anchordoqui:2023tln, Konoplya:2023fmh, Ebadi:2023xhq, Abe:2023yrw, Bari:2023rcw, Ye:2023xyr, Basilakos:2023xof, Jin:2023wri, Bousder:2023ida}.
Among these GW sources, we are particularly interested in domain walls and their link to new physics beyond the standard model (SM).

Domain walls (DWs) are two-dimensional topological defects which could be formed when a discrete symmetry of the scalar potential is spontaneously broken in the early universe~\cite{Kibble:1976sj}.
They are boundaries separating spatial regions with different degenerate vacua.
Stable DWs are thought to be a cosmological problem~\cite{Zeldovich:1974uw}.
As the universe expands, the DW energy density decreases slower than radiation and matter, and would soon dominate the total energy density.
Moreover, large-scale density fluctuations induced by DWs could easily exceed those observed in the cosmic microwave background.

Nonetheless, it is allowed if DWs collapse at a very early epoch~\cite{Vilenkin:1981zs, Gelmini:1988sf, Larsson:1996sp}.
Such unstable DWs can be realized if the discrete symmetry is explicitly broken by a small potential term that gives an energy bias among the minima of the potential.
The bias induces a volume pressure force acting on the DWs that leads to their collapse.
Collapsing DWs significantly produce GWs~\cite{Preskill:1991kd, Gleiser:1998na, Hiramatsu:2010yz, Kawasaki:2011vv}, which would form a stochastic background remaining to the present time, and it could be the one probed by recent PTA experiments.

In this work, we consider a scalar field $S$ with a spontaneously broken $Z_2$-symmetric potential to be the origin of DWs. These DWs can be described by the kink solution of the equation of motion. Since the DWs should collapse before they overclose the universe, a tiny but nonzero $Z_2$-violating potential needs to be added. We will present a possibility that the $Z_2$-violating potential is radiatively originated from a Yukawa interaction between $S$ and a fermionic field $\chi$. The dominant effect comes from one-loop tadpole diagrams for $S$.

Our further analysis shows that the Yukawa coupling should be feeble for reproducing the observed GW data, and this is also required by the freeze-in mechanism of dark matter (DM) production. Therefore, it is possible that the fermion $\chi$, acting as a feebly interacting massive particle (FIMP)~\cite{Hall:2009bx, Bernal:2017kxu}, constitutes the DM relic of the universe. There are a lot of recent studies on exploring such FIMPs~\cite{Hambye:2018dpi,Bian:2018mkl,Belanger:2018sti,No:2019gvl,Brooijmans:2020yij,Dvorkin:2020xga,Calibbi:2021fld,Bian:2021dmp,Elor:2021swj,Ghosh:2022fws,Bhattiprolu:2022sdd, Jiang:2023xdf}. We will explore the interplay between the PTA observations of the SGWB and the freeze-in DM\footnote{Earlier works on the link of GWs from collapsing DWs to feebly interacting DM can be found in Refs.~\cite{Ramazanov:2021eya, Babichev:2021uvl}.}.

The remainder of the paper is outlined as follows.
In Sec.~\ref{sec:DW_GW}, we discuss unstable DWs from the spontaneous breaking of an approximate $Z_2$ symmetry and the resulting GWs.
In Sec.~\ref{sec:DM}, the freeze-in DM production and the induced $Z_2$-violating potential are studied.
In Sec.~\ref{sec:data}, we investigate the parameter ranges simultaneously fulfilling the recent PTA GW data and the observed DM relic density.
Section~\ref{sec:sum} gives a summary.

\section{Domain walls and gravitational waves}
\label{sec:DW_GW}

In this work, we consider the following Lagrangian for scalar fields,
\begin{equation}
\mathcal{L} = \frac{1}{2}\partial_\mu S \partial^\mu S +D_\mu H^\dagger D^\mu H -V_0(H,S),
\end{equation}
where $H$ is the SM Higgs field and $S$ is a real scalar field that is a SM gauge singlet.
The zero-temperature potential $V_0(H,S) = V_{Z_2}+ V_\mathrm{vio}$ consists of the $Z_2$-conserving terms
\begin{equation}\label{eq:V_Z2} 
V_{Z_2} = -\frac{1}{2}\mu_S^2 S^2 + \mu_H^2 |H|^2 +\frac{1}{4}\lambda_S S^4+\lambda_H|H|^4 +\frac{1}{2}\lambda_{HS} |H|^2 S^2,
\end{equation}
which respect a $Z_2$ symmetry $S\to -S$, and the $Z_2$-violating terms
\begin{equation}\label{eq:z2violate}
V_\mathrm{vio} =  \kappa_1S+\frac{\kappa_3}{6}S^3+\kappa_{HS}|H|^2S.
\end{equation}

If one removes the $Z_2$-violating terms, then the Lagrangian has the $Z_2$ symmetry, which would be spontaneously broken for a negative mass parameter $-\mu_S^2$ at low temperatures.
In the phase where both the electroweak and $Z_2$ symmetries are broken, $H$ and $S$ develop nonvanishing vacuum expectation values (VEVs) $\langle H\rangle=(0,v/\sqrt{2})^\mathrm{T}$ and $\langle S\rangle=\pm v_s$ with $v\approx246~\mathrm{GeV}$ and $v_s > 0$.
We assume a hierarchy of $v_s \gg v$, implying that the $Z_2$ symmetry is spontaneously broken at a scale much higher than the electroweak scale.
Furthermore, we assume the Higgs mass parameter $\mu_H^2>0$ and the portal coupling $\lambda_{HS}<0$, and the reason will be explained below.
Thus, when the $S$ field is integrated out below the scale $v_s$, the effective mass parameter for $H$ becomes $\mu_H^2 + \lambda_{HS} v_s^2/2 < 0$, leading to the spontaneous breaking of the electroweak symmetry.
That is to say, the electroweak symmetry breaking is essentially induced by the large VEV of $S$.

At high temperatures, the electroweak and $Z_2$ symmetries would be restored due to thermal corrections to the scalar potential. In the high-temperature limit, the effective potential becomes
\begin{eqnarray}
V_0+V_T(H,S)&=&[\delta m_H^2(T)+\mu_H^2]|H|^2+\frac{1}{2}[\delta m_S^2(T)-\mu_S^2]S^2\nonumber\\
&&+\frac{1}{4}\lambda_S S^4+\lambda_H|H|^4 +\frac{1}{2}\lambda_{HS} |H|^2 S^2+\cdots ,
\end{eqnarray}
where $T$ is the temperature, and $\delta m_H^2(T)$ and $\delta m_S^2(T)$ are thermal corrections to the masses of $H$ and $S$ given by
\begin{eqnarray}
\delta  m_H^2(T) &\approx& \frac{T^2}{4}\left(\frac{1}{4} g^{\prime 2}+\frac{3}{4} g^2+y_t^2+\frac{1}{6} \lambda_{H S}+2 \lambda_H\right),\\
\delta  m_S^2(T) &\approx& \frac{T^2}{4}\left(\frac{2}{3} \lambda_{H S}+\lambda_S\right),
\end{eqnarray}
where $g'$ and $g$ are the $\mathrm{U}(1)_\mathrm{Y}$ and $\mathrm{SU}(2)_\mathrm{L}$ gauge couplings, and $y_t$ is the Yukawa coupling of the top quark.

At a sufficiently high temperature, because of the positive contributions from the thermal masses, both the electroweak and $Z_2$ symmetries are restored.
As the universe expands and cools down, these symmetries become broken at some critical temperatures.
Since these phase transitions happen at an era after reheating, DWs could be produced after the spontaneous breaking of the $Z_2$ symmetry~\cite{Kibble:1976sj}.

A DW corresponds to a kink solution of the equation of motion for the scalar field $S$ given by~\cite{Zeldovich:1974uw}
\begin{equation}
S(z)=v_s\tanh\left(\sqrt{\frac{\lambda_S}{2}}v_sz\right),
\end{equation} 
where the direction perpendicular to the DW is assumed to lie along the $z$-axis.
Thus, $S(z)$ approaches the VEVs $\pm v_s$ for $z \to \pm \infty$.
The DW locates at $z = 0$ with a thickness $\delta \approx (\sqrt{\lambda_S/2}v_s)^{-1}$, separating two domains with $S(z) > 0$ and $S(z) < 0$.
By integrating the energy density along the $z$-direction, the surface energy density of the DW, \textit{i.e.}, its tension, is given by 
\begin{equation}\label{eq:sigma}
\sigma=\frac{4}{3}\sqrt{\frac{\lambda_S}{2}}v_s^3.
\end{equation}
Inside each domain with $S\sim S(\pm\infty)\approx \pm v_s$, we can parametrize $H$ and $S$ as
\begin{equation}
H(x) = \frac{1}{\sqrt{2}} \begin{pmatrix}0\\ v+h(x) \end{pmatrix},\quad S(x) = \pm v_s + s(x).
\end{equation}
where $h(x)$ and $s(x)$ are quantum fields describing the fluctuations above the vacuum.
For $v_s \gg v$, and the masses squared of the scalar bosons $h$ and $s$ are approximately given by
\begin{equation}
m_h^2 \approx 2\lambda_H v^2,\quad m_s^2 \approx 2\lambda_S v_s^2.
\end{equation}
In order to ensure that the two vacua are local minima, the quartic couplings in the $Z_2$-conserving potential should satisfy
\begin{equation}
    \lambda_H > 0,\quad \lambda_S > 0,\quad \lambda_{HS}^2 < 4 \lambda_H \lambda_S.
\end{equation}
Further considering the condition to obtain a negative effective mass parameter for the $H$ field driven by the VEV of $S$, the viable range of $\lambda_{HS}$ expressed by the $h$ boson mass and the VEVs is
\begin{equation}
   - \frac{\sqrt{2\lambda_S}\, m_h}{v} < \lambda_{HS} < - \frac{m_h^2}{v_s^2}.
\end{equation}

Once DWs are created, their tension $\sigma$ acts to straighten them against the friction from the interaction with the cosmic plasma.
If the friction effect is important, the SGWB spectrum induced by DWs could be significantly different from the case without friction~\cite{Nakayama:2016gxi}.
In this work, however, the interactions between DWs and the particles in the thermal bath are mediated by the SM Higgs field and only the sufficiently massive SM particles are relevant.
Such interactions are highly suppressed by the small mixing between $s$ and $h$ because of the hierarchy $v_s \gg v$.
Furthermore, at sufficiently low temperatures\footnote{As we will see below, the temperatures relevant to the annihilation of DWs are of $\sim \mathcal{O}(10^2)~\si{MeV}$.}, the friction force due to the massive SM particles, whose masses arise from the electroweak symmetry breaking, is significantly damped by the exponentially suppressed number densities of these particles~\cite{Saikawa:2017hiv}.

In addition, the Higgs profile inside the DWs is also relevant to friction.
If $\mu_H^2<0$ and $\lambda_{HS}>0$, the SM Higgs boson mass is given by $m_h^2 \approx -(2\mu_H^2+\lambda_{HS}v_s^2) \approx (125~\mathrm{GeV})^2$, implying $|\mu_H| \sim v_s \gg v$ for $\lambda_{HS} \sim \mathcal{O}(1)$.
Inside a DW, the $S$ field value is close to zero, and $\mu_H^2 \sim -v_s^2$ forces the $H$ field to take a large value $\sim v_s$.
Consequently, the SM particles coupled to $H$ become very heavy inside the DW, and their reflection probability with the DW would be highly increased, leading to significant friction~\cite{Nakayama:2016gxi}.
Nonetheless, because we have assumed $\mu_H^2>0$ and $\lambda_{HS}<0$, the $H$ field value would vanish inside the DWs, and the friction can be safely neglected in this study.

Since the friction is negligible, DWs will quickly enter the scaling regime and their energy density evolves as
\begin{align}
\rho_\mathrm{DW} =\frac{\mathcal{A}\sigma}{t},
\end{align}
where $\mathcal{A}\approx 0.8\pm0.1$ is a numerical factor given by lattice simulation~\cite{Hiramatsu:2013qaa}.
$\rho_\mathrm{DW} \propto t^{-1}$ implies that the DW energy density is diluted more slowly than radiation and matter. Therefore, if DWs are stable, they would soon dominate the evolution of the universe, and it conflicts with cosmological observations.
This can be evaded if an explicit $Z_2$-violating potential like Eq.~\eqref{eq:z2violate} presents.

A small $Z_2$-violating potential generates a small energy bias between the two minima of the total potential.
It leads to a volume pressure force acting on the DWs. Thus, the walls could collapse at a very early epoch before they overclose the universe, and would not cause a cosmological problem.
With the $Z_2$-violating potential \eqref{eq:z2violate}, the minima are shifted to
\begin{eqnarray}
v_\pm\approx \pm v_s-\delta, \text{ with }\delta\approx\frac{2\kappa_1+\kappa_3v_s^2}{4\lambda_Sv_s^2},
\end{eqnarray}
where we have neglected the contribution from the $|H|^2S$ term for $v\ll v_s$. We define $\hat{S}(x) \equiv S(x)+\delta$ and rewrite the potential with the redefined scalar field $\hat{S}$:
\begin{eqnarray}\label{eq:V_Shat}
V(\hat{S})=\frac{\lambda_S}{4}(\hat{S}^2-v_s^2)^2+\epsilon v_s\left(\frac{1}{3}\hat{S}^2-v_s^2\right)\hat{S},
\end{eqnarray}
where 
\begin{equation}\label{eq:epsilon}
\epsilon = -\frac{6\kappa_1+\kappa_3v_s^2}{4v_s^3}.
\end{equation}
The two minima of the potential are now located at $\hat{S} = \pm v_s$.

$\hat{S} = +v_s$ corresponds to the true vacuum, while $\hat{S} = -v_s$ corresponds to the false vacuum with slightly higher energy.
The energy difference between them is~\cite{Saikawa:2017hiv}
\begin{eqnarray}\label{eq:Vbias}
V_{\mathrm{bias}}=V(-v_s)-V(+v_s)=\frac{4}{3}\epsilon v_s^4.
\end{eqnarray}
The volume pressure force caused by this energy bias acts on the DWs and tends to make the false vacuum domains shrink.
The collapse of DWs begin when the volume pressure force becomes comparable to the tension force.
As a result, the annihilation temperature of DWs can be estimated by~\cite{Hiramatsu:2013qaa, Saikawa:2017hiv}
\begin{eqnarray}\label{eq:Tann}
T_{\mathrm{ann}}&=&34.1~\mathrm{MeV}~  \mathcal{A}^{-1 / 2} \left[\frac{g_*\left(T_{\mathrm{ann}}\right)}{10}\right]^{-1 / 4}\left(\frac{\sigma}{\mathrm{TeV}^3}\right)^{-1 / 2}\left(\frac{V_{\mathrm{bias}}}{\mathrm{MeV}^4}\right)^{1 / 2}\nonumber\\
&=&76.3~\mathrm{MeV} ~ \mathcal{A}^{-1 / 2}  \left[\frac{g_\ast\left(T_{\mathrm{ann}}\right)}{10}\right]^{-1 / 4}\left(\frac{0.2}{\lambda_S} \frac{m_s}{10^5~\mathrm{GeV}} \frac{\epsilon}{10^{-26}}\right)^{1/2},
\end{eqnarray}
where $g_*$ represents the effective number of relativistic degrees of freedom for the energy density of the plasma and its numerical value depending on the temperature can be found in Ref.~\cite{Husdal:2016haj}.

There are two lower bounds on the energy bias between the two minima $V_{\mathrm{bias}}$~\cite{Saikawa:2017hiv}.
The first one is 
\begin{eqnarray}\label{eq:Vb_lower_bound_1}
V_{\mathrm{bias}}^{1/4}> 0.0218~\mathrm{MeV}~ \mathcal{A}^{1 / 2}\left(\frac{\sigma}{\mathrm{TeV}^3}\right)^{1 / 2},
\end{eqnarray}
given by the requirement that DWs should collapse before they dominate the universe.
Moreover, the energetic particles produced from DW collapse could destroy the light elements generated in the big bang nucleosynthesis (BBN).
Thus, we should require that DWs annihilate before the BBN epoch.
This leads to a second lower bound as
\begin{eqnarray}
V_{\mathrm{bias}}^{1/4}> 0.507~\mathrm{MeV}~ \mathcal{A}^{1 / 4}\left(\frac{\sigma}{\mathrm{TeV}^3}\right)^{1 / 4}.
\end{eqnarray}

The stochastic GWs from collapsing DWs can be estimated by numerical simulations~\cite{Hiramatsu:2010yz, Kawasaki:2011vv, Hiramatsu:2013qaa}.
The frequency spectrum of the SGWB is commonly characterized by
\begin{equation}
\Omega_\mathrm{GW}(f) = \frac{f}{\rho_\mathrm{c}} \frac{d\rho_\mathrm{GW}}{d f},
\end{equation}
where $\rho_\mathrm{GW}$ is the GW energy density and $\rho_\mathrm{c}$ is the critical energy density of the universe.
At high frequencies, the simulations show that the GW spectrum behaves as $\Omega_{\mathrm{GW}} \propto f^{-1}$.
At small frequencies, the spectrum scales as $\Omega_{\mathrm{GW}} \propto f^3$ because of causality~\cite{Caprini:2009fx, Hiramatsu:2013qaa}.
The peak of the spectrum at the DW annihilation temperature $T_\mathrm{ann}$ can be expressed as~\cite{Hiramatsu:2013qaa}
\begin{equation}
\Omega_\mathrm{GW}^\mathrm{peak}\Big|_{T=T_{\mathrm{ann}}} = \frac{3 \tilde{\epsilon}_\mathrm{GW} \alpha_*^2}{32\pi},
\end{equation}
where $\tilde{\epsilon}_\mathrm{GW} = 0.7 \pm 0.4$ is derived from numerical simulation.
$\alpha_\ast$ represents the ratio of the GW energy density to the radiation energy density $\rho_\mathrm{rad}$ at $T_\mathrm{ann}$, \textit{i.e.},
\begin{equation}\label{eq:alpha}
\alpha_\ast\equiv\left.\frac{\rho_{\mathrm{DW}}}{\rho_{\mathrm{rad}}}\right|_{T=T_{\mathrm{ann}}}=0.035 \left[\frac{10}{g_\ast(T_{\mathrm{ann}})}\right]^{1/2}\frac{\mathcal{A}}{0.8}\frac{0.2}{\lambda_S}\left(\frac{m_s}{10^{5}~\mathrm{GeV}}\right)^3\left(\frac{100~\mathrm{MeV}}{T_{\mathrm{ann}}}\right)^2.
\end{equation}

Taking into account the dilution of the GW energy density due to the cosmological expansion, the peak amplitude of the SGWB spectrum at the present time can be expressed as~\cite{Hiramatsu:2013qaa,Saikawa:2017hiv}
\begin{eqnarray}
\Omega_{\mathrm{GW}}^{\mathrm{peak}}h^2 &=& 7.2 \times 10^{-18}~ \tilde{\epsilon}_{\mathrm{GW}} \mathcal{A}^2\left[\frac{g_{\ast s}\left(T_{\mathrm{ann}}\right)}{10}\right]^{-4 / 3}\left(\frac{\sigma}{1~ \mathrm{TeV}^3}\right)^2\left(\frac{T_{\mathrm{ann}}}{10~\mathrm{MeV}}\right)^{-4}\nonumber\\
&=& 5.9\times 10^{-9}\tilde{\epsilon}_{\mathrm{GW}} \mathcal{A}^4 \left[\frac{g_{\ast s}\left(T_{\mathrm{ann}}\right)}{10}\right]^{-4 / 3}\frac{g_{\ast}\left(T_{\mathrm{ann}}\right)}{10}\left(\frac{10^{-26}}{\epsilon}\right)^{2}\left(\frac{m_s}{10^5~\mathrm{GeV}}\right)^4,
\label{eq:Omega_peak}
\end{eqnarray}
where $g_{\ast s}$ denotes the effective number of relativistic degrees of freedom for the entropy density.
The present GW peak frequency can be estimated by the Hubble rate at $T_\mathrm{ann}$ taking into account the redshift effect, given by
\begin{eqnarray}
f_{\mathrm{peak}} &=& 1.1 \times 10^{-9} ~\mathrm{Hz}~\left[\frac{g_*\left(T_{\mathrm{ann}}\right)}{10}\right]^{1 / 2}\left[\frac{g_{* s}\left(T_{\mathrm{ann}}\right)}{10}\right]^{-1 / 3}\frac{T_{\mathrm{ann}}}{10~\mathrm{MeV}}\nonumber\\
&=& 8.39\times10^{-9}\mathrm{~Hz}~\mathcal{A}^{-1/2} \left[\frac{g_{\ast s}\left(T_{\mathrm{ann}}\right)}{10}\right]^{-1/3}\left[\frac{g_{\ast}\left(T_{\mathrm{ann}}\right)}{10}\right]^{1/4}
\left(\frac{0.2}{\lambda_S} \frac{m_s}{10^5~\mathrm{GeV}} \frac{\epsilon}{10^{-26}}\right)^{1/2}.\qquad
\end{eqnarray}
Thus, the present SGWB spectrum induced by collapsing DWs can be evaluated by
\begin{equation}
\Omega_{\mathrm{GW}}(f) h^2 = \Omega_{\mathrm{GW}}^{\mathrm{peak}} h^2 \times \begin{cases}
 \left(\dfrac{f}{f_\mathrm{peak}}\right)^3, & f < f_\mathrm{peak},\\[1em]
\dfrac{f_\mathrm{peak}}{f}, & f > f_\mathrm{peak}.
\end{cases}
\end{equation}

\begin{figure}[!t]
\centering	
\includegraphics[width=0.55\textwidth]{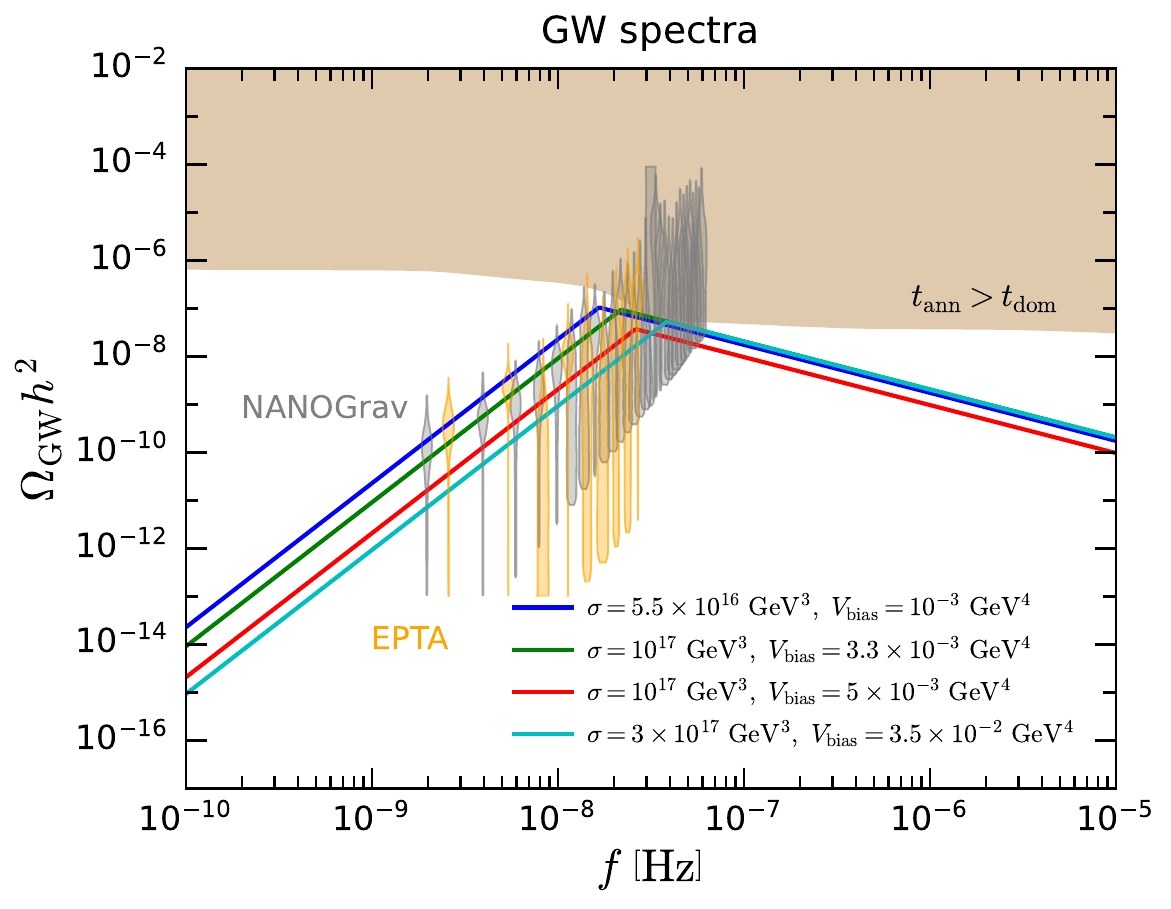}
\caption{GW spectra generated by collapsing DWs for four sets of parameters.
The gray and orange violins represent the reconstructed posterior distributions for the NANOGrav (gray violins)~\cite{NANOGrav:2023hvm} and EPTA (orange violins)~\cite{Antoniadis:2023zhi} observations of nHz GWs, respectively.
The brown region is excluded by the requirement that DWs should annihilate before they dominate the universe.}
\label{fig:DW_GW_spectra}
\end{figure}

In Fig.~\ref{fig:DW_GW_spectra}, we show the GW spectra generated by DWs for some benchmark parameters, compared with the reconstructed posterior distributions for the NANOGrav (gray violins)~\cite{NANOGrav:2023hvm} and EPTA (orange violins)~\cite{Antoniadis:2023zhi} signals.
We find that the spectra with $\sigma \sim \mathcal{O}(10^{17})~\mathrm{GeV}^3$ and $V_\mathrm{bias} \sim \mathcal{O}(10^{-3})~\mathrm{GeV}^4$ can explain the PTA observations. 
As discussed above, DWs must annihilate before they dominate the energy of the universe.
Thus, the time $t_\mathrm{ann}$ when DWs annihilate should be earlier than the time $t_\mathrm{dom}$ when DWs would dominate.
According to Eqs.~\eqref{eq:Tann}, \eqref{eq:Vb_lower_bound_1} and \eqref{eq:Omega_peak}, this gives upper limits on the GW spectra from collapsing DWs. In Fig.~\ref{fig:DW_GW_spectra}, the unavailable region corresponding to $t_\mathrm{ann} > t_\mathrm{dom}$ is shaded by the brown color.

\section{Freeze-in dark matter and the induced $Z_2$-violating potential}
\label{sec:DM}

So far, the $Z_2$-violating potential is introduced by hand.
In the following, we will consider it to be generated by loops of fermionic dark matter through a Yukawa interaction with the scalar field $S$.
To be precise, we introduce a Dirac fermion field $\chi$, which is a singlet under all the SM gauge symmetries. The Lagrangian involving $\chi$ is
\begin{equation}
\mathcal{L}_\chi=\bar{\chi}(i\slashed{\partial} - m_\chi)\chi + y_\chi S\bar{\chi}\chi,
\end{equation}
where $y_\chi$ is the Yukawa coupling constant.
When $S$ acquires a nonzero VEV, $\langle S\rangle \approx \pm v_s$, the mass of $\chi$ receives a correction, $m^{(\pm)}_\chi \approx m_\chi\mp y_\chi v_s$.
In this work, we assume that $m_\chi\gg y_\chi v_s$, so $m^{(\pm)}_\chi\approx m_\chi$ holds.

After reheating, $s$ bosons are in thermal equilibrium with the SM particles due to the $|H|^2S^2$ interaction, while $\chi$ fermions are assumed to be out of equilibrium with nearly vanishing number density.
This requires $y_\chi$ to be a feeble coupling constant.
In this case, $\chi$ fermions could be produced via $s$ decays but never reach thermal equilibrium if $y_\chi$ is extremely small, say, $\sim\mathcal{O}(10^{-10})$.
This is the well-known freeze-in mechanism of DM production~\cite{Hall:2009bx}, and $\chi$ acts as a DM candidate.
The evolution of the DM number density $n_\chi$ is determined by the Boltzmann equation~\cite{Belanger:2018ccd}
\begin{equation}\label{eq:boltz}
\frac{d n_\chi}{dt}+3Hn_\chi\approx \frac{m_S^2 T}{\pi^2}  \Gamma_{s \rightarrow \chi\bar{\chi}} \tilde{K}_1\left(x_S, 0,0, 1, 0,0\right),
\end{equation}
where $x_i\equiv m_i/T$, and $\Gamma_{s \rightarrow \chi\bar{\chi}}$ is the $s \rightarrow \chi\bar{\chi}$ partial decay width given by
\begin{equation}
\Gamma_{s \rightarrow \chi\bar{\chi}}\approx\frac{y_\chi^2m_s}{8\pi}
\end{equation}
for $m_\chi\ll m_s$.
The function $\tilde{K}_1\left(x_1, x_2, x_3, \eta_1, \eta_2, \eta_3\right) $ is defined as
\begin{equation}
\tilde{K}_1\left(x_1, x_2, x_3, \eta_1, \eta_2, \eta_3\right) \equiv \frac{1}{(4 \pi)^2 p_{\mathrm{CM}} T} \int \prod_{i=1}^3\left(\frac{d^3 p_i}{E_i} \frac{1}{e^{E_i / T}-\eta_i}\right) e^{E_1 / T} \delta^{(4)}\left(p_1-p_2-p_3\right)
\end{equation}

By solving Eq.~\eqref{eq:boltz}, the $\chi$ number density at the present time $t_0$ can be approximated by~\cite{Belanger:2018ccd}
\begin{equation}
n_\chi(t_0) \approx \frac{3.434 s_0 M_{\mathrm{Pl}}\Gamma_{s\rightarrow \chi\bar{\chi}}}{[g_\ast(m_s/3)]^{3/2}m_s^2},
\end{equation}
where $g_\ast(m_s/3)\approx 108$, $s_0$ is the present entropy density, and $M_{\mathrm{Pl}}\approx 2.4\times10^{18}~\mathrm{GeV}$ is the reduced Planck mass.
The $\chi$ relic density is then given by
\begin{equation}
\Omega_\chi h^2\approx 2.74\times10^{8}~\frac{m_\chi}{\mathrm{GeV}} \frac{n_\chi(t_0)}{s_0} \approx 8.13\times10^{22}~\frac{y_\chi^2 m_\chi}{m_s}.
\end{equation}
On the other hand, the observed value of the DM relic density is $\Omega_{\mathrm{DM}} h^2 = 0.1200\pm 0.0012$~\cite{Planck:2018vyg}, which implies that the Yukawa coupling $y_\chi$ should be feeble.

In the potential \eqref{eq:V_Shat}, the $Z_2$-violating term is characterized by the parameter $\epsilon$, which is related to $\kappa_1$ and $\kappa_3$ via Eq.~\eqref{eq:epsilon}.
Taking $\lambda_S \sim 0.2$, $\sigma \sim 10^{17}~\mathrm{GeV}^3$, and $V_\mathrm{bias} \sim 10^{-3}~\mathrm{GeV}^4$, which lead to a GW spectrum accounting for the recent PTA observations, we obtain $\epsilon \sim \mathcal{O}(10^{-26})$ from Eqs.~\eqref{eq:sigma} and \eqref{eq:Vbias}.
Note that the $S\bar{\chi}\chi$ Yukawa interaction explicitly breaks the $Z_2$ symmetry even if the tree-level $Z_2$-violating potential is absent.
It is natural to conjecture that such an extremely tiny $\epsilon$ is originated from the feeble Yukawa interaction through $\chi$ loops.

\begin{figure}[!t]
\centering	
\includegraphics[width=0.3\textwidth]{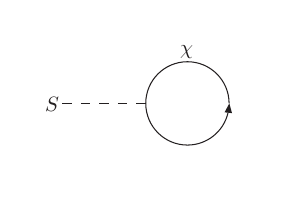}
\includegraphics[width=0.3\textwidth]{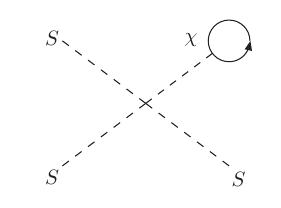}
\includegraphics[width=0.3\textwidth]{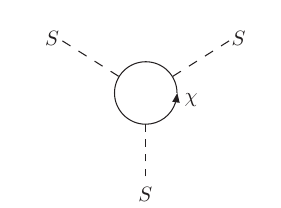}
\caption{Examples of one-loop diagrams for generating the $Z_2$-violating potential terms.}
\label{fig:loop_diag}
\end{figure}

In Fig.~\ref{fig:loop_diag}, we show some one-loop diagrams relevant to the generation of 
the $Z_2$-violating couplings $\kappa_1$ and $\kappa_3$.
The first two diagrams contains the one-loop tadpole diagrams of $S$ and they give the dominant contributions to $\kappa_1$ and $\kappa_3$. Although the third diagram also contributes to $\kappa_3$, it is negligible compared with the second diagram due to a $y_\chi^3$ suppression. Our further calculation leads to the $\epsilon$ value at the $m_s$ scale as
\begin{eqnarray}\label{eq:loop_induced_epsilon}
\epsilon(m_s)\approx\frac{3\lambda_S^{3/2}y_\chi}{\sqrt{2}\pi^2}\left(\frac{m_\chi}{m_s}\right)^3\ln\frac{\Lambda_{\mathrm{UV}}}{m_s},
\end{eqnarray}
where $\epsilon = 0$ at some ultraviolet (UV) scale $\Lambda_{\mathrm{UV}}$ is assumed.
Below we assume $\Lambda_{\mathrm{UV}}=M_{\mathrm{Pl}}$.
Thus, $\lambda_S \sim 0.2$ and $v_s\sim 10^5~\mathrm{GeV}$ lead to $\epsilon(m_s)\sim 0.6y_\chi(m_\chi/m_s)^3$.

\begin{figure}[!t]
	\centering
        \includegraphics[width=0.55\textwidth]{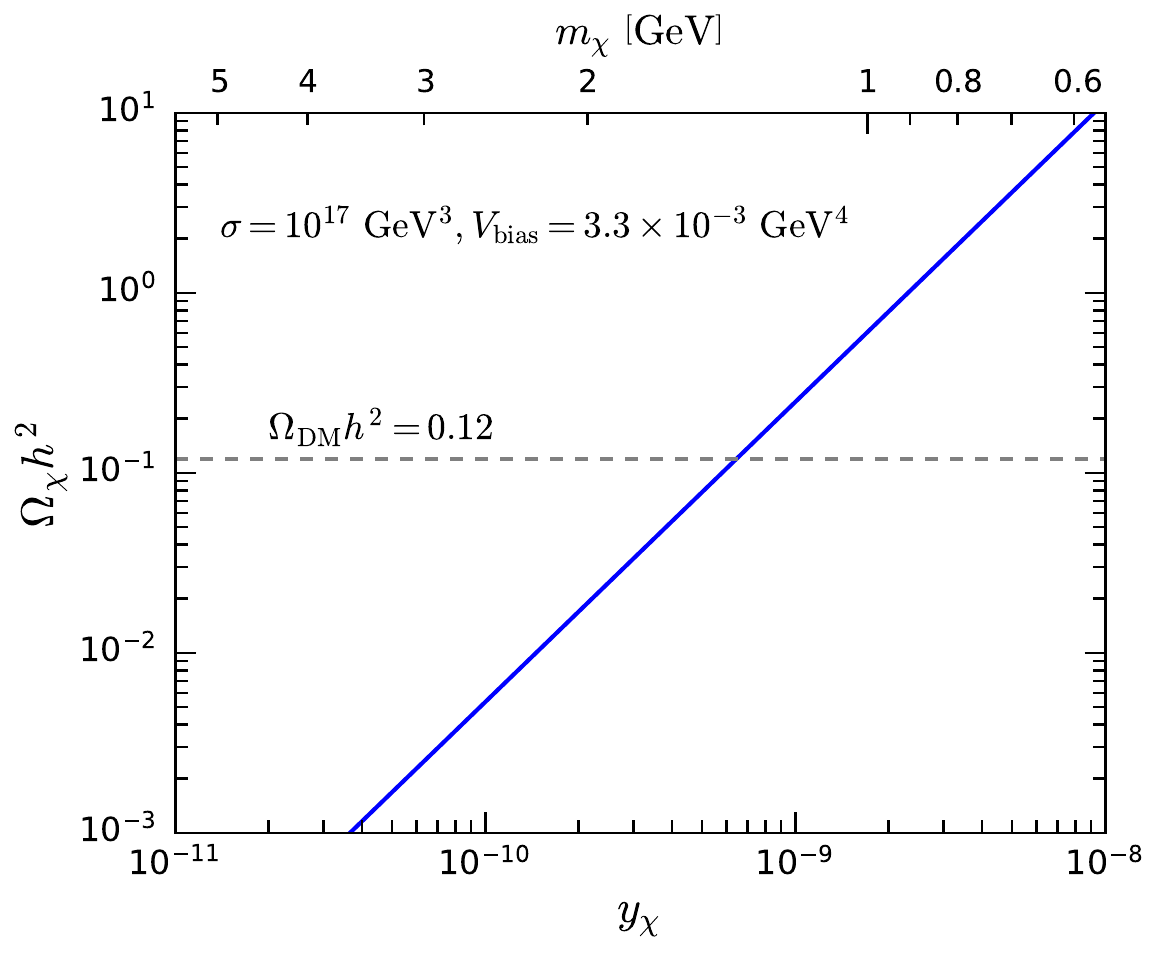}
	\caption{DM relic density as a function of the Yukawa coupling $y_\chi$ with fixed $\sigma=10^{17}~\mathrm{GeV}^3$, $V_\mathrm{bias}=3.3\times10^{-3}~\mathrm{GeV}^4$, and $\lambda_S = 0.2$. The upper horizontal axis denotes the value of $m_\chi$. The horizontal dashed line corresponds to the Planck observation value of the DM relic density~\cite{Planck:2018vyg}.}
	\label{fig:DM_relic_density}
\end{figure}

By taking $\lambda_S = 0.2$, $\sigma=10^{17}~\mathrm{GeV}^3$, and $V_\mathrm{bias}=3.3\times10^{-3}~\mathrm{GeV}^4$, which correspond to the GW spectrum denoted by the green line in Fig.~\ref{fig:DW_GW_spectra}, we obtain $v_s = 6.19\times 10^5~\mathrm{GeV}$, $m_s = 3.91\times 10^5~\mathrm{GeV}$, 
$\epsilon = 3.58\times 10^{-26}$, 
$T_\mathrm{ann} = 163~\mathrm{MeV}$, $\Omega_\mathrm{GW}^\mathrm{peak}h^2 = 9.44\times 10^{-8}$, and $f_\mathrm{peak} = 2.18\times 10^{-8}~\mathrm{Hz}$.
Then, $y_\chi$ and $m_\chi$ are related by Eq.~\eqref{eq:loop_induced_epsilon}.
For this set of parameters, the predicted DM relic density as a function of $y_\chi$ is shown in Fig.~\ref{fig:DM_relic_density}, with the upper horizontal axis corresponding to $m_\chi$.
We find that both the extremely tiny $\epsilon \sim \mathcal{O}(10^{-26})$ and the observed DM relic density $\Omega_\mathrm{DM}h^2=0.12$ can be naturally explained by the feeble Yukawa coupling $y_\chi \sim \mathcal{O}(10^{-10})$.
Therefore, our theory can simultaneously explain the recent PTA observations of nHz GWs and the DM relic via freeze-in production.

\section{Favored parameter ranges}
\label{sec:data}

In this section, we investigate the parameter ranges favored by both the PTA GW observations and the observed DM relic density.
Our model has four free parameters, which can be chosen to be $\lambda_S$, $y_\chi$, $m_s$, and $m_\chi$.
In the following analysis, we fix the quartic coupling $\lambda_S=0.2$ to reduce one free parameter.

\begin{figure}[!t]
\centering	
\includegraphics[width=0.55\textwidth]{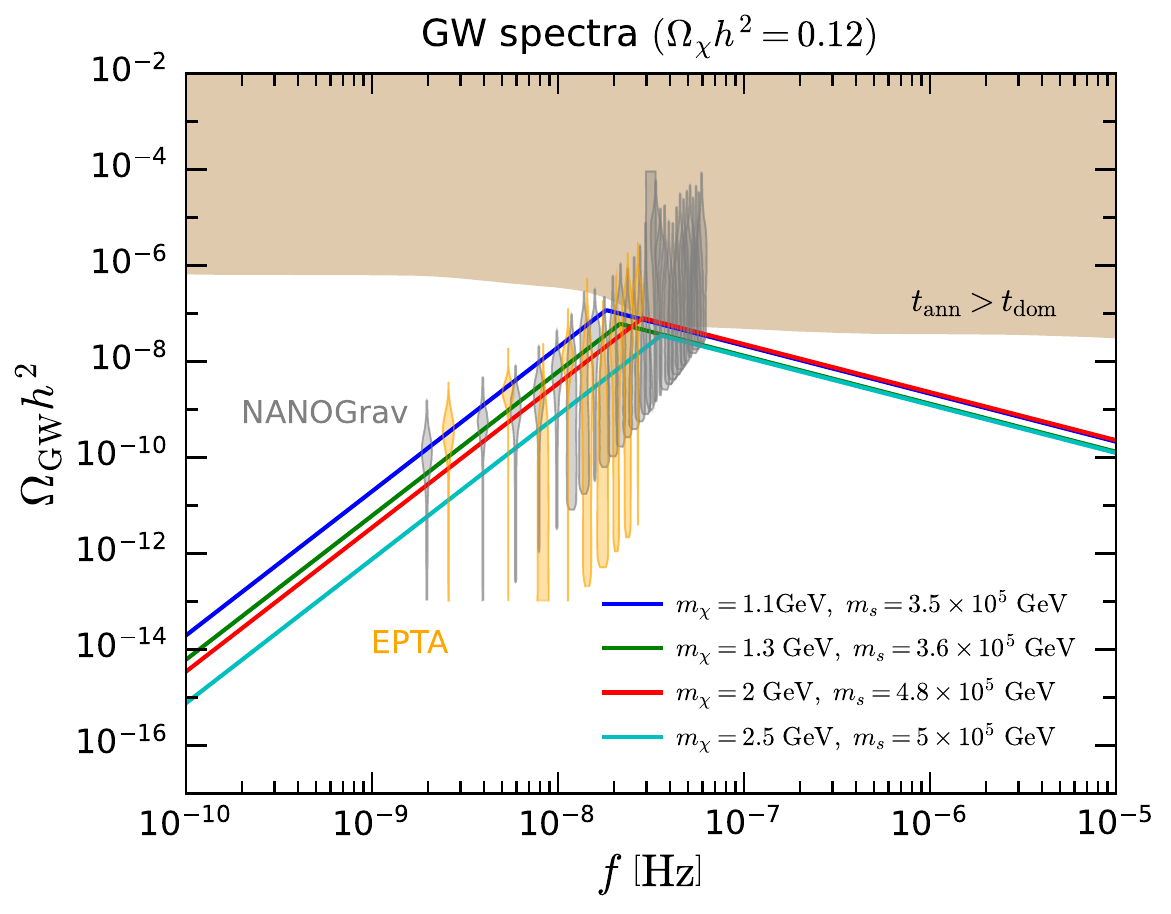}
\caption{Same as in Fig.~\ref{fig:DW_GW_spectra}, but for GW spectra corresponding to different values of $m_\chi$ and $m_s$ with $\lambda_S = 0.2$ fixed and $y_\chi$ adjusted to give $\Omega_\chi h^2 = 0.12$.}
\label{fig:FIDM_GW_spectra}
\end{figure}

In Fig.~\ref{fig:FIDM_GW_spectra}, the GW spectra for four benchmark points with $m_\chi = 1.1\text{--}2.5~\mathrm{GeV}$ and $m_s = (3.5\text{--}5)\times 10^5~\mathrm{GeV}$ are shown.
For all these benchmark points, the Yukawa coupling $y_\chi$ is adjusted to give the mean value of the observed DM relic density $\Omega_\chi h^2 = 0.12$.
We can see that the GW spectrum is quite sensitive to $m_\chi$ and $m_s$, limiting them varying within roughly one order of magnitude.

\begin{figure}[!t]
\centering	
\includegraphics[width=0.55\textwidth]{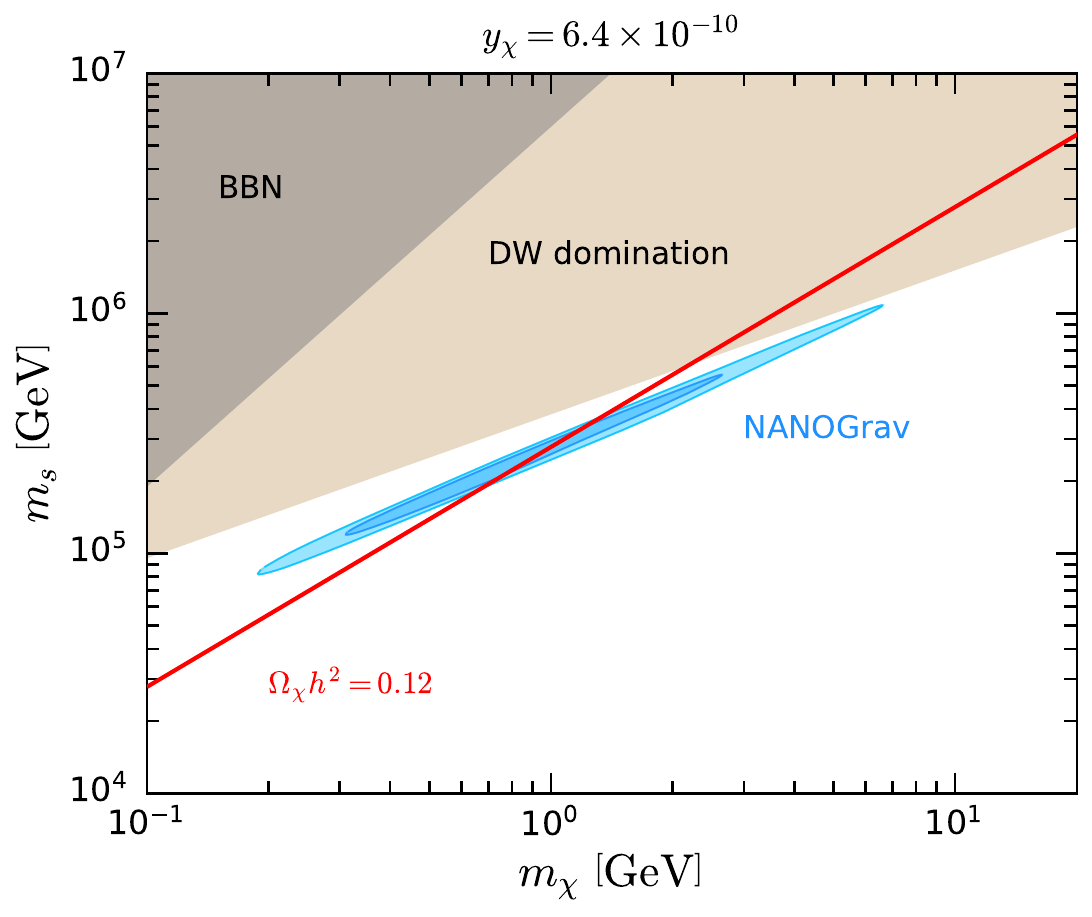}
\caption{Parameter regions favored by the NANOGrav GW signal in the $m_\chi$-$m_s$ plane for $y_\chi=6.4 \times 10^{-10}$.
The deep blue and light blue regions corresponds to the 68\% and 95\% Bayesian credible regions favored by the NANOGrav data, respectively.
The red line denotes the mean value of the Planck observation of the DM relic density, $\Omega_\chi h^2 = 0.12$.
The brown and gray regions are excluded because DWs would dominate the universe and would inject energetic particles to affect BBN, respectively.}
\label{fig:NANOGrav_fit_FIDM_1}
\end{figure}

In our model, after the annihilation of DWs, most of their energy releases to the SM thermal bath.
In this case, the NANOGrav collaboration has reconstructed the posterior distributions of $(T_\mathrm{ann},\alpha_\ast)$ accounting for the observed nHz GW signal, as shown in the left panel of Fig.~12 in Ref.~\cite{NANOGrav:2023hvm}.
We use this result to study the favored parameter regions.
In Fig.~\ref{fig:NANOGrav_fit_FIDM_1}, the Yukawa coupling is fixed as $y_\chi=6.4\times10^{-10}$, and the deep blue and light blue regions in the $m_\chi$-$m_S$ plane correspond to the 68\% and 95\% Bayesian credible regions favored by the NANOGrav data.
It shows a high correlation between $m_\chi$ and $m_S$, which can be understood by the behavior of the GW peak amplitude, $\Omega_\mathrm{GW}^\mathrm{peak} h^2 \propto m_s^{10} / m_\chi^{6}$, inferred from Eqs.~\eqref{eq:Omega_peak} and \eqref{eq:loop_induced_epsilon}.

In Fig.~\ref{fig:NANOGrav_fit_FIDM_1}, $\Omega_\chi h^2 = 0.12$ corresponds to a red line, which intersects the regions favored by the NANOGrav GW observation.
Thus, both the PTA GW signal and the observed DM relic density can be well interpreted.
Moreover, the brown and gray regions are  excluded by the requirements that DWs should collapse before they dominate the universe and before the BBN epoch, respectively.
We find that these constraints do not exclude the NANOGrav 95\% Bayesian credible region.

\begin{figure}[!t]
	\centering
	\subfigure[$y_\chi=4.6\times10^{-10}$.]
	{\includegraphics[width=0.48\textwidth]{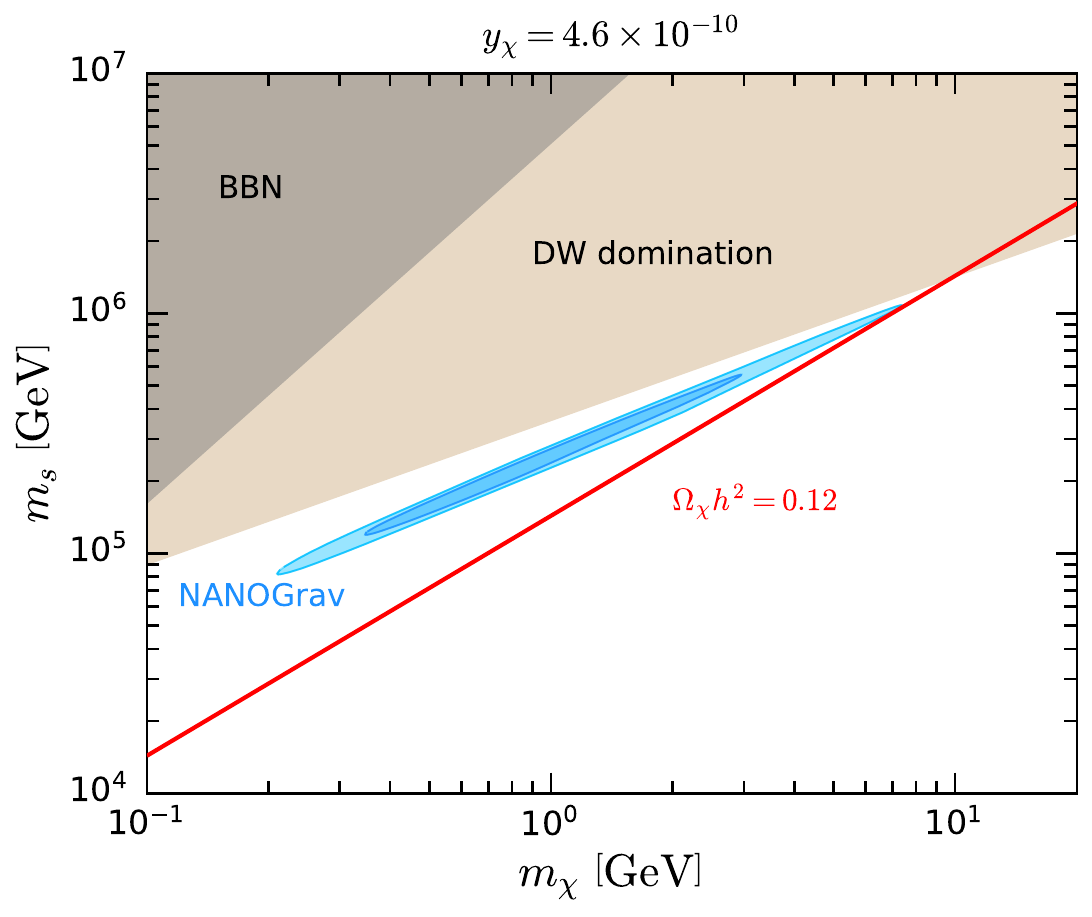}}
        \hspace{.01\textwidth}
        \subfigure[$y_\chi=8.7\times10^{-10}$.]
	{\includegraphics[width=0.48\textwidth]{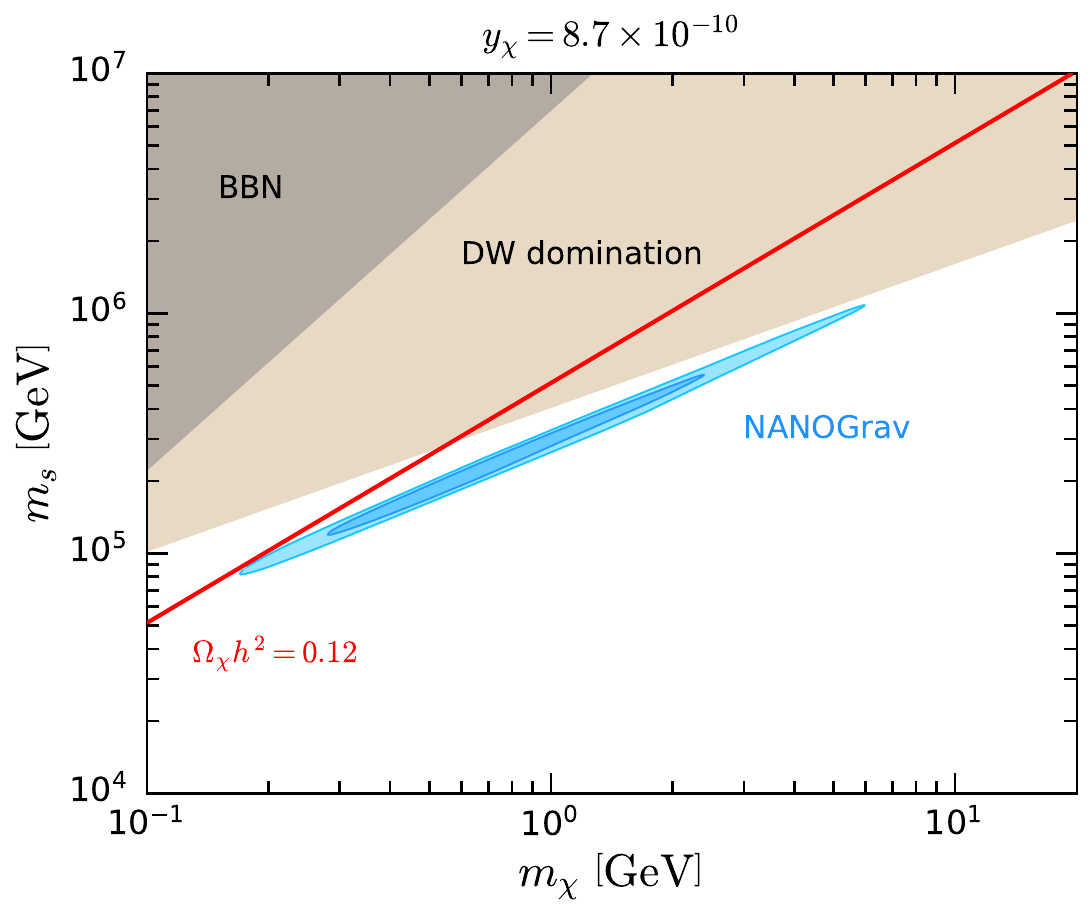}}
	\caption{Same as in Fig.~\ref{fig:NANOGrav_fit_FIDM_1}, but for $y_\chi=4.6\times10^{-10}$ (a) and $y_\chi=8.7\times10^{-10}$ (b).}
	\label{fig:NANOGrav_fit_FIDM_2_3}
\end{figure}

The intersection of the $\Omega_\chi h^2 = 0.12$ line and the NANOGrav favored regions sensitively depends on the value of $y_\chi$.
Our calculation shows that such a intersection can only happen for $4.6\times 10^{-10} \lesssim y_\chi \lesssim 8.7\times 10^{-10}$.
In the left and right panels of Fig.~\ref{fig:NANOGrav_fit_FIDM_2_3}, we demonstrate the results for $y_\chi = 4.6\times 10^{-10}$ and $y_\chi = 8.7\times 10^{-10}$, respectively.
In both cases, the $\Omega_\chi h^2 = 0.12$ line can only touch the edge of the NANOGrav 95\% Bayesian credible region.
To sum up, for $\lambda_S = 0.2$, we find that the preferred parameter ranges where our model can simultaneously explain the NANOGrav GW signal and the DM relic density are
\begin{equation}
\begin{aligned}
&4.6\times10^{-10} \lesssim y_\chi \lesssim 8.7\times10^{-10},
\\
&0.17~\mathrm{GeV} \lesssim m_\chi\lesssim 7.5~\mathrm{GeV},
\\
&8.1\times10^4~\mathrm{GeV} \lesssim m_s \lesssim 10^6~\mathrm{GeV}.
\end{aligned}
\end{equation}

\section{Summary}
\label{sec:sum}

In this work, we studied the interplay between the SGWB from collapsing DWs and FIMP DM.
The newly introduced real scalar field $S$ satisfies a $Z_2$ symmetry in the tree-level potential, which, however, is violated by the Yukawa coupling $y_\chi$ with a fermion field $\chi$.
The linear and cubic terms of $S$ can be induced by the Yukawa coupling at one-loop level, and they explicitly break the $Z_2$ symmetry of the potential, leading to an energy bias between the two minima.
Thus, after the spontaneous breaking of the $Z_2$ symmetry, unstable DWs would be formed.
We considered that the Yukawa coupling is feeble, \textit{i.e.}, $y_\chi \sim \mathcal{O}(10^{-10})$, and the $\chi$ fermions become FIMPs that are produced by the freeze-in mechanism, accounting for dark matter in the universe.

On the other hand, four PTA collaborations have recently reported evidences of a SGWB at nHz frequencies, which could be produced by the collapse of DWs in the early universe.
Since the tiny $Z_2$-violating potential induced by the feeble Yukawa coupling leads to unstable DWs, it is possible that our model can explain the PTA data.
Comparing with the posterior distributions in the GW spectrum reconstructed by the NANOGrav and EPTA data, our analysis showed that the $Z_2$-violating coefficient should be as tiny as $\epsilon \sim 10^{-26}$, which can be naturally induced by the feeble Yukawa couplings $y_\chi \sim \mathcal{O}(10^{-10})$ at one-loop level.
Thus, our scenario is very suitable for interpreting the PTA observations of the nHz SGWB.

Moreover, we investigated the parameter regions where both the PTA GW observations and the DM relic density can be simultaneously explained.
We found that  the parameters should satisfy $y_\chi \in (4.6\times10^{-10},8.7\times10^{-10})$, $m_\chi \in (0.17, 7.5)~\mathrm{GeV}$, and $m_s \in (8.1\times10^4,10^6)~\mathrm{GeV}$ for a fixed quartic scalar couplings $\lambda_S = 0.2$.
The corresponding regions also fulfill the requirements that DWs should collapse before they overclose the universe and they should not affect the BBN.

\begin{acknowledgments}

This work is supported by the National Natural Science Foundation of China (NSFC) under Grants No.~12275367, No.~11905300, and No.~11875327, the Fundamental Research Funds for the Central Universities, the Guangzhou Science and Technology Planning Project under Grant No.~2023A04J0008, and the Sun Yat-Sen University Science Foundation.

\end{acknowledgments}

\bibliographystyle{utphys}
\bibliography{ref}

\end{document}